\documentclass[aps,pra,twocolumn,amsmath,amssymb,showpacs]{revtex4-1}
\usepackage{graphicx}
\usepackage{amsmath}
\usepackage{bm}
\usepackage{cleveref}

\newcommand{\be}{\begin{equation}}  
\newcommand{\ee}{\end{equation}}
\newcommand{\ba}{\begin{array}}
\newcommand{\ea}{\end{array}}
\newcommand{\bea}{\begin{eqnarray}}
\newcommand{\eea}{\end{eqnarray}}

\newcommand{\bra}{\langle}
\newcommand{\ket}{\rangle}

\newcommand{\nn}{\nonumber}
\newcommand{\ca}{{\cal A}}

\begin{document}

\title{Heat flow reversals without reversing the arrow of time: the role of internal quantum coherences and correlations} 

\author{C.L. Latune$^{1,2}$, I. Sinayskiy$^{1}$, F. Petruccione$^{1,2,3}$}
\affiliation{$^1$Quantum Research Group, School of Chemistry and Physics, University of
KwaZulu-Natal, Durban, KwaZulu-Natal, 4001, South Africa\\
$^2$National Institute for Theoretical Physics (NITheP), KwaZulu-Natal, 4001, South Africa\\
$^3$School of Electrical Engineering, KAIST, Daejeon, 34141, Republic of Korea}

\date{\today}
\begin{abstract}
One of the stunning consequences of quantum correlations in thermodynamics is the reversal of the arrow of time, recently shown experimentally in [K. Micadei, et al., Nat. Commun. {\bf 10}:2456 (2019)], and manifesting itself by a reversal of the heat flow (from the cold system to the hot one). Here, we show that contrary to what could have been expected, heat flow reversal can happen without reversal of the arrow of time. Moreover, contrasting with previous studies, no initial correlations between system and bath is required. Instead, the heat flow reversal only relies on {\it internal} quantum coherences or correlations, which provides practical advantages over previous schemes: one does not need to have access to the bath in order to reverse the heat flow. The underlying mechanism is explained and shown to stem from the collective system-bath coupling and the impact of non-energetic coherences (coherences between degenerate energy levels) on apparent temperatures. The phenomenon is first uncovered in a broad framework valid for diverse quantum systems containing energy degeneracy. By the end of the paper, aiming at experimental realisations, more quantitative results are provided for a pair of two-level systems. Finally, as a curiosity, we mention that our scheme can be adapted as a correlations-to-energy converter, which have the particularity to be able to operate at constant entropy, similarly to ideal work sources.
 \end{abstract}

\maketitle

\section{Introduction}
Recently, Micadei et al. shown in a two-level atom-based experiment \cite{Micadei_2017} the reversal of the arrow of time thanks to initial correlations between system and reservoir. The macroscopic manifestation of the reversal of the arrow of time is an astonishing heat flow reversal: from the coldest system to the hottest one.
Along with the work of Micadei \& al., a series of theoretical studies on reversal of the arrow of time \cite{Partovi_2008, Jennings_2010, Jevtic_2012, Jevtic_2015, Henao_2018,Levi_2019} illustrates one of the most striking consequence of correlations between quantum systems. 
On a broader perspective, recent years have seen a huge effort towards understanding the role and impact of quantum effects such as correlations and coherences on thermodynamics tasks as work extraction \cite{Oppenheim_2002,Scully_2003,Hovhannisyan_2013,Llobet_2015,Aberg_2014}, thermal machines \cite{Dillenschneider_2009,Scully_2011,Uzdin_2015,Niedenzu_2015,Altintas_2015,Uzdin_2016,Chen_2016,Barrios_2017}, quantum battery charging \cite{Binder_2015,Ferraro_2018}, quantum transport \cite{Caruso_2009,Robentrost_2009,Ishizaki_2009,Lloyd_2011,Lee_2017}, and natural or synthetic light harvesting systems \cite{Scully_2010,Svidzinsky_2011,Creatore_2013,Romero_2014,Lloyd_2011,Collini_2010,Huelga_2013,Chin_2013}.

Here, we report one more surprising effect of correlations and coherences: heat flows reversal {\it without} reversing the arrow of time. One additional major difference with the previous works \cite{Micadei_2017,Partovi_2008, Jennings_2010, Jevtic_2012, Jevtic_2015,Henao_2018,Levi_2019} is that in our configuration there is no need of initial correlations between the system and reservoir, but only coherences (or correlations) {\it within} the system (which can be composed of several subsystems). 
This should simplify its experimental verification and increase its practical interest since it means that the heat flow can be reversed by acting only on the system instead of acting on both the system and reservoir.
Moreover, the reported effect is shown to be valid for a large class of systems, the essential ingredients being energy degeneracy and collective coupling. We use the framework introduced in \cite{paperapptemp} on the concept of apparent temperature and the role of coherences and correlations in heat flows. The results relevant for this paper are recalled in Section \ref{secindisting}.
We also provide an alternative viewpoint using mutual information (Section \ref{central}), representing an interesting link with previous studies \cite{Micadei_2017,Jennings_2010}. 
Further in the paper we illustrate the heat flow reversal with a pair of two-level systems (one of the the simplest and experimentally accessible system containing energy degeneracy). 
While the study in \cite{bathinducedcohTLS} focused on the energetic and entropic consequences of bath-induced coherences in a pair of two-level systems, the illustration presented in this paper shows how coherences {\it initially present} within the pair can lead to curious dynamics such as heat flow reversal. 
 Finally, the scheme can alternatively be used as an intriguing correlations-to-energy converter, which can be tuned to operate at constant entropy, reproducing ideal work source or external (classical) power source \cite{Kosloff_2013,Kosloff_2014}.
 As a side note, we mention a study reporting also heat flow reversals, but in a different context of topological insulators and topological effects \cite {Rivas_2017}.

\section{Indistinguishability and apparent temperature}\label{secindisting}
We consider a system $S$ interacting with a thermal bath $B$ at temperature $T_B$ (and inverse temperature denoted by $\beta_B=1/T_B$, $k_B\equiv 1$). 
We assume that $S$ can be described by a single energy transition $\omega$, and that $S$ contains energy degeneracy, which includes, but not limited to, atoms with degenerate energy levels as for instance three-level atoms (sometimes referred to as $\Lambda$ and $V$ energy configurations) \cite{Scully_Book}, ensembles of two-level systems, ensembles of spins of arbitrary size, and ensembles of harmonic oscillators. Then, the free Hamiltonian of $S$ has the form $H_S=\sum_{n}\sum_{i=1}^{l_n} \epsilon_n |n,i\ket\bra n,i|$, where $|n,i\ket$ is an eigenvector associated to the eigenenergy $\epsilon_n$. The index $i$ running from 1 to $l_n\geq 1$ denotes the degeneracy of the energy level $n$.  
We consider a coupling between $S$ and the bath of the form
\be\label{coupling}
V = g A_S B
\ee
where $g$ corresponds to the coupling strength, $A_S$ is an observable of $S$ and $B$ and observable of the bath. The energy transitions of $S$ involved in the coupling can be obtained from $A_S$ by $\Pi_n A_S \Pi_{n'}$, where $\Pi_n$ is the projector onto the eigenspace associated to the eigenenergy $\epsilon_n$, which can be expressed in terms of the eigenvectors as 
\be\label{projector}
\Pi_n=\sum_{i=1}^{l_n} |n,i\ket\bra n,i|.
\ee 

The form of the coupling \eqref{coupling} implies  that all the transitions $\Pi_n A_S \Pi_{n'}$ ``see'' the same bath. Conversely, the transitions are {\it indistinguishable} from the point of view of the bath: an absorption of a bath excitation can activate indiscernibly any resonant transition. We mention this point to emphasise that in general such indistinguishability requires some experimental engineering like parallel transition dipole moments \cite{ Tscherbul_2015} (for atomic systems) or subsystems (if $S$ is an ensemble of subsystems) at spatial locations which are indistinguishable, or indiscernible, from the bath. This last point can usually be obtained by confinement in a volume much smaller than the typical variation length scale of the bath (as for instance in superradiance \cite{Gross_1982,Devoe_1996,Barnes_2005}) or by adding an ancillary system between the $S$ and the bath to erase part of the information ``seen'' by the bath as in \cite{Wood_2014,Wood_2016,Hama_2018,Niedenzu_2018} and experimentally realised for instance in \cite{Barberena_2019, Raimond_1982, Moi_1983, Devoe_1996, Barnes_2005, Goban_2015, Guerin_2016, Araujo_2016, Weiss_2019}. 

All transitions $\Pi_n A_S \Pi_{n'}$ of same energy $\nu=\epsilon_{n'}-\epsilon_n$ can be put together to form the eigenoperators (or ladder operators) \cite{Petruccione_Book} associated to the observable $A_S$, 
\be\label{eigenoperator}
{\cal A}(\nu)=\sum_{\epsilon_{n'}-\epsilon_n=\nu}  \Pi_n A_S \Pi_{n'}.
\ee
 The observable $A_S$ can be re-written as a sum of its eigenoperators, $A_S=\sum_{\nu} {\cal A}(\nu)$. Since we assumed that $S$ is a single energy transition system (or at least only a single energy transition couples to the bath), the only eigenoperators different from zero are for $\nu=\pm\omega$. For simplicity, in the remainder of the paper we use the notation, ${\cal A}(\nu=\omega) \equiv {\cal A}$ and ${\cal A}(\nu=-\omega)\equiv {\cal A}^{\dag}$. We recall that the eigenoperators satisfy the following commutation relation $[H_S,{\cal A}]=-\omega {\cal A}$ and $[H_S,{\cal A}^{\dag}]=\omega {\cal A}^{\dag}$. Note that when $S$ is an ensemble of $n$ subsystems $S_i$ indistinguishable from the bath, the eigenoperators $\ca$ and $\ca^{\dag}$ are the sum of the local eigenoperator $a_i$ and $a_i^{\dag}$ of each subsystem $S_i$, $\ca = \sum_{i=1}^n a_i$ and $\ca^{\dag} =\sum_{i=1}^n a_i^{\dag}$. In this context, the coupling \eqref{coupling} is a {\it collective} coupling between the subsystems $S_i$ and the bath. By contrast, an {\it independent} coupling of the subsystems with the bath would correspond to a situation where each subsystem $S_i$ is coupled to an independent bath operator $B_i$ (or even an independent bath).


 Under weak coupling between $S$ and the bath, the Markov and Born approximations are valid \cite{Petruccione_Book, Cohen_Book} and the reduced dynamics of $S$ is given by a Markovian master equation (see Appendix \ref{appheatflow}). The direction of the heat flow between $S$ and $B$ is given by the sign of energy variation of $S$, $\dot E_S:={\rm Tr}\dot\rho_S H_S$. Using the master equation describing the dynamics of $S$ one obtains $\dot E_S \propto \big(e^{-\omega/T_B}-e^{-\omega/{\cal T}_S}\big)$ (see Appendix \ref{appheatflow}), where ${\cal T}_S$ is defined by \cite{paperapptemp} ($\hbar=1$, $k_B=1$)
\be\label{apptemp}
{\cal T}_S := \omega \left(\ln\frac{\bra {\cal A} {\cal A}^{\dag} \ket}{\bra {\cal A}^{\dag} {\cal A}\ket}\right)^{-1},
\ee
with $\bra {\cal O} \ket :={\rm Tr}\rho_S {\cal O}$ denoting the expectation value of the operator ${\cal O}$ evaluated in the state $\rho_S$. The parameter ${\cal T}_S$ is called the apparent temperature of $S$ since it determines the direction of the heat flow: $\dot E_S >0$ ($\dot E_S <0$) if and only if ${\cal T}_S<T_B$ (${\cal T}_S>T_B$). 
Importantly, in the situation where $S$ is an ensemble of subsystems, this result remains valid even if the subsystems are interacting between each other (as long as the interactions conserve the total energy of the ensemble). 

Crucially, the apparent temperature ${\cal T}_S$ takes into account coherences contained in $S$ \cite{paperapptemp}. 
 This can be seen simply by injecting the expressions of the eigenoperators \eqref{eigenoperator} and of the projectors \eqref{projector} in the expectation values $\bra {\cal A} {\cal A}^{\dag} \ket$ and $\bra {\cal A}^{\dag} {\cal A} \ket$. This yields 
\bea
\bra {\cal A} {\cal A}^{\dag} \ket&=&\sum_{\epsilon_{n'}-\epsilon_n=\omega}\sum_{i}^{l_n} \bra n,i|A_S\Pi_{n'}A_S|n,i\ket \bra n,i|\rho|n,i\ket\label{diagterm}\\
&+&\sum_{\epsilon_{n'}-\epsilon_n=\omega}\sum_{i\ne i'}^{l_n} \bra n,i|A_S\Pi_{n'}A_S|n,i'\ket \bra n,i'|\rho|n,i\ket,\nn\\ \label{offdiagterm}
\eea
where the first term \eqref{diagterm} contains the contribution from the populations $\bra n,i|\rho|n,i\ket$ of $S$ and the second term \eqref{offdiagterm} contains the contributions from the coherences $\bra n,i'|\rho|n,i\ket$ between {\it degenerate} levels. In the remainder of the paper  such coherences are called non-energetic coherences as opposed to coherences between non-degenerate levels (levels of different energy) called energetic coherences. 
Note that the terms in \eqref{offdiagterm} are non-zero if $A_S$ contains degenerate transitions (from one level $|n,i\ket$ to two or more degenerate levels $|m,j\ket$, $|m,j'\ket$), 
which is the essence of 
 the collective or indistinguishable coupling \eqref{coupling} described above. 
  Interestingly, 
 no energetic coherence contributes to the expectation values $\bra \ca \ca^{\dag}\ket$. The expression of $\bra\ca^{\dag}\ca\ket$ can be obtained directly from \eqref{diagterm} and \eqref{offdiagterm} by substituting $\omega$ by $-\omega$.



It is important to keep in mind that non-energetic coherences in a many-body system can take the form of correlations between subsystems \cite{paperapptemp}. For instance, in the simple situation of a pair of two-level systems considered in Section \ref{secTLS}, the term $\chi^0 =  \alpha |0\ket|1\ket \bra 1|\bra0| + \alpha^{*}|1\ket|0\ket\bra0|\bra1|$ in \eqref{corrtermTLS} represents a correlation between the two subsystems (since it implies $\rho^0_S\ne\rho_{S_1}^0\rho_{S_2}^0$) but it also corresponds to (non-energetic) coherences between the degenerate states $|01\ket$ and $|10\ket$.

Due to the presence of the term \eqref{offdiagterm} in the expectation value of $\ca\ca^{\dag}$ and $\ca^{\dag}\ca$, it is possible to manipulate the apparent temperature of $S$ only by introducing non-energetic coherences within $S$. Moreover, when $S$ is a many-body system, correlations between subsystems can correspond to non-energetic coherences (as just mentioned above) and therefore also affect the apparent temperature ${\cal T}_S$. In other words, correlated subsystems in an ensemble have an apparent temperature which can largely differ from the apparent temperature of a non-correlated ensemble but otherwise identical (meaning same local state of each subsystem). In particular, when $S$ is in a thermal state at temperature $T_S < T_B$ (each subsystem is in a thermal state at temperature $T_S$), generating correlations between the subsystems can make the resulting apparent temperature ${\cal T}_S$ larger than $T_B$. This suggests that a cold ensemble interacting with a hot bath can appear indeed hotter than the bath and be refrigerated by this hot bath thanks to initial correlations between the subsystems, referred to as {\it internal correlations} in the following. Conversely, a hot ensemble can be further heated up by interacting with a cold bath thanks to initial internal correlations. We investigate these curious phenomena in the following.

\section{Reversing the heat flow}\label{sechfreversal}
Formalising the above ideas, we consider in this section a many-body system $S$ initially in a state
\be\label{initialstt}
\rho^0_S = \rho^{\rm th}_S(\beta_S) + \chi^0
\ee
composed of a thermal contribution $\rho^{\rm th}_S(\beta_S):=Z^{-1}(\beta_S)e^{-\beta_SH_S}$ of temperature $T_S=1/\beta_S$ and partition function $Z(\beta_S):={\rm Tr}e^{-\beta_SH_S}$, upgraded by the term $\chi^0$ containing arbitrary non-energetic coherences in form of correlations. 
Importantly, correlations disappear upon partial trace so that each subsystem of $S$ is {\it locally in a thermal state at temperature $T_S$}. Therefore, assuming for instance that the bath temperature $T_B$ is larger than $T_S$ one expects a heat flow from the bath to $S$. This intuitive view omits the role of the correlations between the subsystems. Indeed, the resulting apparent temperature of $S$, given by \eqref{apptemp}, can be decomposed in the following form 
\be\label{corrcontribution}
\frac{\omega}{{\cal T}_S}= \frac{\omega}{T_S} +    \log \frac{1+ c^{+}}{1+c^{-}},
\ee
with $c^{-}:= \bra \ca^{\dag} \ca \ket_{\rm cor}/\bra \ca^{\dag} \ca \ket_{\rm loc}$ and $c^{+}:=\bra \ca \ca^{\dag} \ket_{\rm cor}/\bra \ca \ca^{\dag} \ket_{\rm loc}$. The expectation value $\bra \ca \ca^{\dag} \ket$ has been split into a local contribution $\bra \ca \ca^{\dag} \ket_{\rm loc}:={\rm Tr} \rho^{\rm th}(\beta_S)\ca\ca^{\dag}$ corresponding to the term \eqref{diagterm}, and a contribution from the correlations $\bra \ca \ca^{\dag} \ket_{\rm cor}:={\rm Tr} \chi^0\ca\ca^{\dag}$ corresponding to the term \eqref{offdiagterm}. A similar splitting was made for $\bra \ca^{\dag} \ca \ket$. Equation \eqref{corrcontribution} shows that the contributions from the correlations add up to the inverse temperature $\beta_S$ (corresponding to the local contribution) which can result in an apparent temperature ${\cal T}_S$ larger than $T_B$, implying a {\it heat flow reversal}, from $S$ to the bath. The necessary and sufficient conditions for heat flow reversal are
\be\label{condoncor1}
{\cal C} > \bra \ca^ {\dag} \ca \ket_{\rm loc} \frac{e^{\omega \beta_S} - e^{\omega \beta_B}}{e^{\omega \beta_B}-1} >0
\ee
where ${\cal C} :=\bra \ca^{\dag} \ca \ket_{\rm cor} = \bra \ca \ca^{\dag} \ket_{\rm cor}$ (the equality being a direct consequence of $\chi^0$ representing correlations between subsystems, see Appendix \ref{equalityofcor}), and $\beta_B=1/T_B$ is the inverse bath temperature. In other words, the inequality \eqref{condoncor1} establishes the minimal conditions on the correlations within the ensemble $S$ in order to have a reversal of the heat flow. 
Since ${\cal C}$ is typically limited by a value of the order of $\bra \ca^ {\dag} \ca \ket_{\rm loc}$ (to ensure the positivity of $\rho^0$), one can see from \eqref{condoncor1} that the heat flow reversal is in general possible only for a limited range of inverse temperatures $\beta_S$ around $\beta_B$. Moreover, as expected, the closer $\beta_S$ and $\beta_B$, the weaker is the condition on the correlations. 

Conversely, if $T_S>T_B$, one would expect a heat flow from $S$ to the bath, but again, the contribution from the correlations can {\it reverse} the heat flow. This happens if and only if 
\be\label{condoncor2}
{\cal C} <   \bra \ca^ {\dag} \ca \ket_{\rm loc} \frac{e^{\omega \beta_S} - e^{\omega \beta_B}}{e^{\omega \beta_B}-1} <0.
\ee
As above, $|{\cal C}|$ is limited by a value of the order of $\bra \ca^ {\dag} \ca \ket_{\rm loc}$ so that the heat flow reversal can happen only for a limited range of inverse temperature $\beta_S$ around $\beta_B$. 

The above considerations can be extended to negative temperatures for $S$ but also for the bath. Indeed, effective baths at negative temperatures (emerging for instance from spin baths \cite{Kosloff_2019,Assis_2018}, compositions of thermal baths at different positive temperatures \cite{Brunner_2012}, or in a context of thermal machines \cite{autonomousmachines}) are quite common in thermodynamics and plays an important role.
 Then, when $\beta_B$ is negative, heat flow reversals are still possible but the inequality signs in conditions \eqref{condoncor1} and \eqref{condoncor2} are inverted. 

One other interesting extension of the above picture is considering arbitrary initial state. Then, the thermal state $\rho^{\rm th}_S(\beta_S)$ in the decomposition \eqref{initialstt} of the initial state $\rho_S^0$ is substituted by an arbitrary product of local states of each subsystem of $S$. 
 The above results remain valid by substituting the initial temperature $T_S$ by the apparent temperature of the product of the local states. 

Additionally, one should keep in mind that one underlying necessary ingredient for the heat flow reversal is the {\it collective} coupling \eqref{coupling} introduced in Section \ref{secindisting}. Indeed, when each subsystem interacts with an independent local bath, initial correlations between subsystems cannot be ``seen'' by local baths so that each subsystem thermalises to the thermal equilibrium state at the local bath temperature (see also Section \ref{secillustration} and Fig. \ref{enVStime}).

Finally, for the sake of completeness, as suggested in the previous Section \ref{secindisting}, heat flow reversals can be achieved with single system containing non-energetic coherences. Considering an initial state of the form \eqref{initialstt} with $\chi^0$ containing non-energetic coherences and assuming that $T_B>T_S$ (of arbitrary signs), the conditions for heat flow reversals can be obtained in a similar way as for many-body systems, leading to ${\cal C}^{+}e^{\omega\beta_B}-{\cal C}^{-}>\bra {\cal A}^{\dag}{\cal A}\ket_{\rm loc} \left(e^{\omega\beta_S}-e^{\omega\beta_B}\right)>0$, 
where ${\cal C}^{+} :=  \bra {\cal A}{\cal A}^{\dag}\ket_{\rm coh} \ne {\cal C}^{-} :=  \bra {\cal A}^{\dag}{\cal A}\ket_{\rm coh}$, with the convention $\bra {\cal O}\ket_{\rm coh}:= {\rm Tr}\chi^0{\cal O}$, for any operator ${\cal O}$. When $T_B<T_S$, non-energetic coherences leading to heat flow reversal must satisfy ${\cal C}^{+}e^{\omega\beta_B}-{\cal C}^{-}<\bra {\cal A}^{\dag}{\cal A}\ket_{\rm loc} \left(e^{\omega\beta_S}-e^{\omega\beta_B}\right)<0$.

In the next section, focusing on many-body systems, we analyse in detail how this heat flow reversal is related to the one reported in \cite{Micadei_2017}. Interestingly, although both phenomena can be described within the same formalism, they are different in nature. In particular, the present heat flow reversal is not related to reversal of the arrow of time by contrast to \cite{Micadei_2017}.

\section{Decrease of mutual information}\label{central}
The reversal of the arrow of time reported in \cite{Micadei_2017} relies on the {\it decrease} of the mutual information between $S$ and the bath $B$, defined by \cite{Nielsen_Book} 
\be
I(S:B) :=  S_{S}+S_B - S_{SB},
\ee
 where $S_X := -{\rm Tr}_X \rho_X \ln \rho_X$, denotes the von Neumann entropy and $\rho_X$ the density operator of the system $X=S, B, SB$, respectively. Since the mutual information is always positive, being equal to zero only when $S$ and $B$ are uncorrelated \cite{Nielsen_Book}, a decrease of mutual information $I(S:B)$ is possible if and only if $S$ and $B$ are initially correlated. As a consequence, the entropy production can become negative \cite{Micadei_2017,Partovi_2008, Jennings_2010, Jevtic_2012,Henao_2018}, corresponding to reversing the arrow of time. One of its surprising macroscopic manifestation is a heat flow reversal \cite{Jennings_2010,Micadei_2017,Henao_2018}. 

Using the same formalism as in \cite{Micadei_2017, Jennings_2010}, we show that even with $S$ and $B$ initially uncorrelated ($I(S:B)=0$), one can still have a heat flow reversal. This also provides an alternative point of view on the results of the previous section \ref{sechfreversal}. The key idea is to substitute the mutual bipartite information $I(S:B)$ by the three partite mutual information $I(S_1:S_2:B)$ defined as follow,
\be
I(S_1:S_2:B) := S_{S_1} + S_{S_2}+S_B - S_{SB}.
\ee
Due to the subadditivity of the entropy \cite{Nielsen_Book}, we have $I(S_1:S_2:B)\geq I(S:B)$ which guarantees the positivity of $I(S_1:S_2:B)$. Then, even if $S$ and $B$ are initially uncorrelated we can still have $I(S_1:S_2:B)>0$ thanks to initial correlations between $S_1$ and $S_2$. This provides a ``fuel'' sufficient to reverse the heat flow as we show in the following.

We derive an expression of $Q$, the heat (exchanged between the initial and final instant of time) from $B$ to $S$, in term of the variation of the mutual information. 
We assume that $S$ is initialised in a state of the form \eqref{initialstt}, that is a product of local thermal states upgraded by some correlations $\chi^0$. Since $B$, $S_1$, and $S_2$ are initially in thermal states, the relative entropy between the initial state $\rho_X^0$ and the state $\rho_X^t$ at an arbitrary instant of time $t$ is equal to
\bea\label{ent}
S(\rho_X^t||\rho_X^0) &:=& {\rm Tr}_X \rho_X^t (\log\rho_X^t -\log\rho_X^0)\nn\\
&=&-S_X + \beta_X E_X^t +\log Z_X,
\eea
where $X$ stands for $S_1$,$S_2$, or $B$, and $E_X^t := {\rm Tr}_X \rho_X^t H_X$. From such identity and since $S(\rho_X^0||\rho_X^0) =0$ one obtains
\be\label{relentr}
S(\rho_X^t||\rho_X^0) = -\Delta S_X +\beta_i \Delta E_X,
\ee
where $\Delta {\cal O} := {\cal O}^t - {\cal O}^0$ is the variation of the quantity ${\cal O}= E_X, S_X$. Finally, assuming that the first law is satisfied (conservation of energy) $\Delta E_S = - \Delta E_B := Q$, an expression of the heat exchanged $Q$ can be obtained as follows
\bea\label{heatflow}
(\beta_S -\beta_B)Q &&=\beta_S \Delta E_S + \beta_B \Delta E_B \nn\\
&&= \beta_S \Delta E_{S_1} + \beta_S \Delta E_{S_2} + \beta_B \Delta E_B \nn\\
&& = \Delta I(S_1:S_2:B) + S(\rho_{S_1}^t||\rho_{S_1}^0) \nn\\ 
&&+ S(\rho_{S_2}^t||\rho_{S_2}^0) + S(\rho_{B}^t||\rho_{B}^0),
\eea
where the identity $\Delta E_S = \Delta E_{S_1}+ \Delta E_{S_2}$ holds exactly if we assume that $S_1$ and $S_2$ are not interacting. 
When $S_1$, $S_2$, and $B$ are initially uncorrelated ($I(S_1:S_2:B)=0$), all the terms in the right-hand side of \eqref{heatflow} are positive so that Eq. \eqref{heatflow} expresses the ``natural'' heat flow, from the hottest to the coldest system.
However, when $S_1$ and $S_2$ are initially correlated, the variation of the tripartite mutual information can be re-written as $\Delta I(S_1:S_2:B) = \Delta I(S_1:S_2) + \Delta S(\rho_B)$, which can become negative. For $S_1$ and $S_2$ highly correlated, the negative contribution from $\Delta I(S_1:S_2)$ can even dominate all other positive contributions in the right-hand side of \eqref{heatflow}, imposing a reversal of the heat flow (when the left-hand side becomes negative). 
This requires initial correlations between $S_1$ and $S_2$ to be high enough, which is expressed by Eqs. \eqref{condoncor1} and \eqref{condoncor2}.
%
%

Importantly, the entropy production, equal to \cite{Esposito_2010,Deffner_2011} $\Sigma = \Delta I(S:B) + S(\rho_B^t|\rho_B^0)$ is always positive, so that the arrow of time is not reversed by contrast with \cite{Micadei_2017}. 

\section{Pair of two-level systems}\label{secTLS}
In this section we focus on the same system considered in \cite{Micadei_2017}, a pair of two-level systems, in order to obtain simple quantitative results which could be verified experimentally. 
As mentioned in the introduction, the thermodynamic effects of {\it bath-induced coherences} in this same system was discussed in \cite{bathinducedcohTLS}. Here, we focus on a different and somehow opposite aspect: how correlations (or coherences) initially present in the system can dramatically affect both its on going evolution and steady state energy, and lead to heat flow reversal. 
Following the previous sections, we assume that the pair of two-level systems interacts with the bath at inverse temperature $\beta_B$ through the collective ladder operators $S^{+} :=\sigma_1^{+} + \sigma_2^{+}$ and $S^{-}:=\sigma_1^{-} + \sigma_2^{-}$, well-known from superradiance \cite{Gross_1982} and playing the role of $\ca^{\dag}$ and $\ca$. The local ladder operators can be expressed in term of the ground and excited state $|0\ket$ and $|1\ket$ of each two-level systems, namely $\sigma_i^{+}:=|1\ket\bra 0|$ and $\sigma_i^{-} = |0\ket\bra 1|$. 
We assume that the pair is initially in a state $\rho_S^0=\rho^{\rm th}_S(\beta_S) +\chi^0$ of the form \eqref{initialstt}, with $H_S=\omega \sum_{i=1}^2\sigma_i^{+}\sigma_i^{-}$, $Z(\beta_S)=(1+e^{-\omega\beta_S})^2$, and 
\be\label{corrtermTLS}
\chi^0 =  \alpha |0\ket|1\ket \bra 1|\bra0| + \alpha^{*}|1\ket|0\ket\bra0|\bra1|
\ee
 (adopting the convention that the tensor product order is taken to be the same for ``bras'' and ``kets''). Note that we choose $\chi^0$ containing only non-energetic coherences since, as mentioned in Section \ref{sechfreversal}, energetic coherences do not play any role in the apparent temperature. This choice also corresponds to the initial correlations in \cite{Micadei_2017}. 

Applying the results of Section \ref{sechfreversal} we have ${\cal C} = {\rm Tr} \chi^0 S^{+}S^{-} = 2\Re \alpha$ ($\Re$ denotes the real part) and $\bra S^{+}S^{-}\ket_{\rm loc} = {\rm Tr} \rho_S^{\rm th}(\beta_S) S^{+}S^{-}=2(1+e^{\omega\beta_S})^{-1}$, so that a reversal of the heat flow happens if and only if
\be\label{realpha1}
\Re \alpha > \alpha_c:=\frac{e^{\omega \beta_S} - e^{\omega \beta_B}}{(e^{\omega\beta_S}+1)(e^{\omega \beta_B}-1)} >0
\ee
for $\beta_S/\beta_B>1$ (remaining valid for temperatures of arbitrary sign).
Conversely, for $\beta_S/\beta_B<1$, the heat flow reversal happens for 
\be\label{realpha2}
\Re \alpha < \alpha_c <0.
\ee

 As mention in Section \ref{sechfreversal}, the authorised values of $\Re \alpha$ are limited by the positivity condition of $\rho_S^0$ which imposes here $|\alpha| \leq e^{-\omega\beta_S}/Z(\beta_B)$, so that $|\Re \alpha| \leq e^{-\omega\beta_S}/Z(\beta_B)$. This yields some constraints on the respective values of $\beta_S$ and $\beta_B$ for which the heat flow can be reversed by internal correlations. For instance considering a bath at positive temperature, the heat flow can be reverted for $T_S<T_B$ only if $T_S$ is not too small compared to $T_B$, in proportions stated by the following inequality,
\be\label{appearinghot}
\omega\beta_B \geq \omega\beta_S- \log{\frac{1+2e^{\omega\beta_S}}{2+e^{\omega\beta_S}}}.
\ee
One should note that the term $\log{\frac{1+2e^{\omega\beta_S}}{2+e^{\omega\beta_S}}}$ is always positive, taking values in the interval $[0;\log 2]$. 
Conversely, if $T_S>T_B>0$, the heat flow can be reversed only if 
\be\label{appearingcold}
\omega\beta_B\leq2\omega\beta_S + \log(1+2e^{-\omega\beta_S}).
\ee
Note that \eqref{appearingcold} is less restrictive than \eqref{appearinghot}. This surprising asymmetry 
means that it is in general more difficult to increase the apparent temperature from correlations than to reduce it. 

\subsection{Illustration}\label{secillustration}
\begin{figure}
\centering
(a)\includegraphics[width=7cm, height=4.5cm]{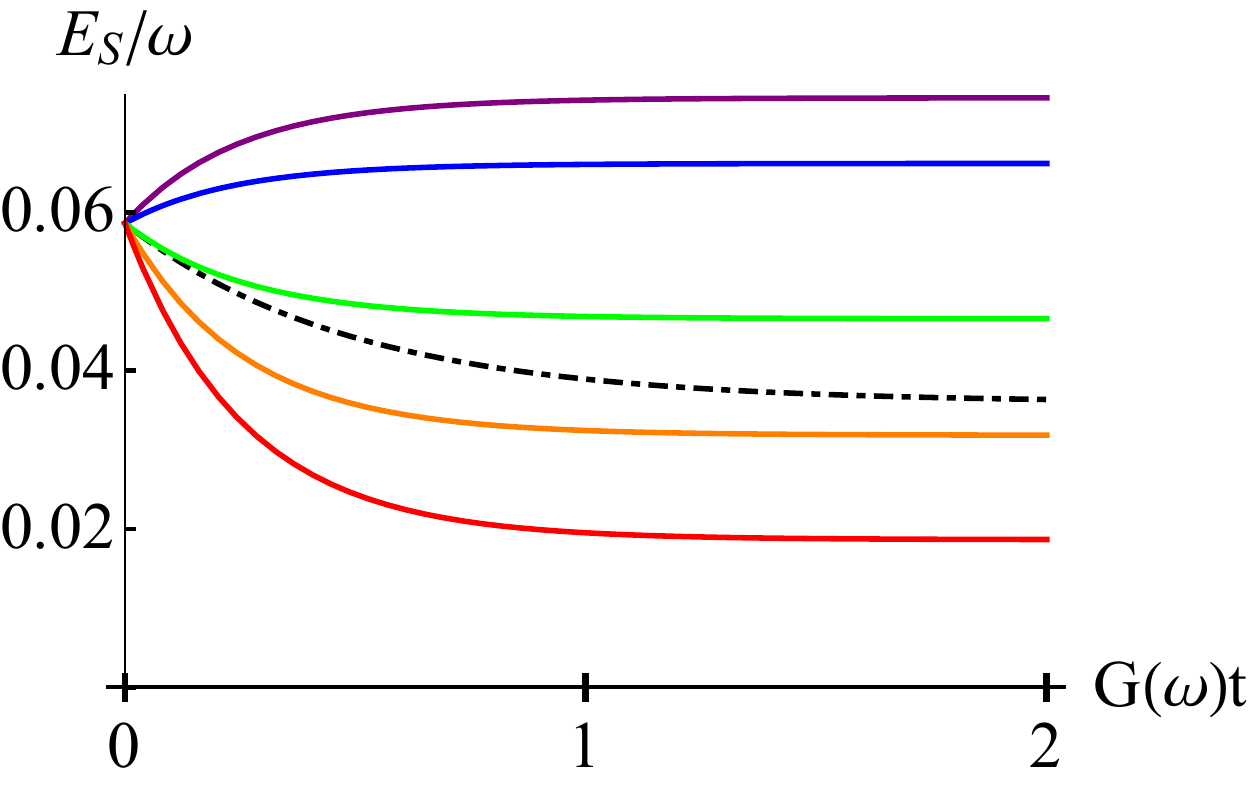}
(b)\includegraphics[width=7cm, height=4.5cm]{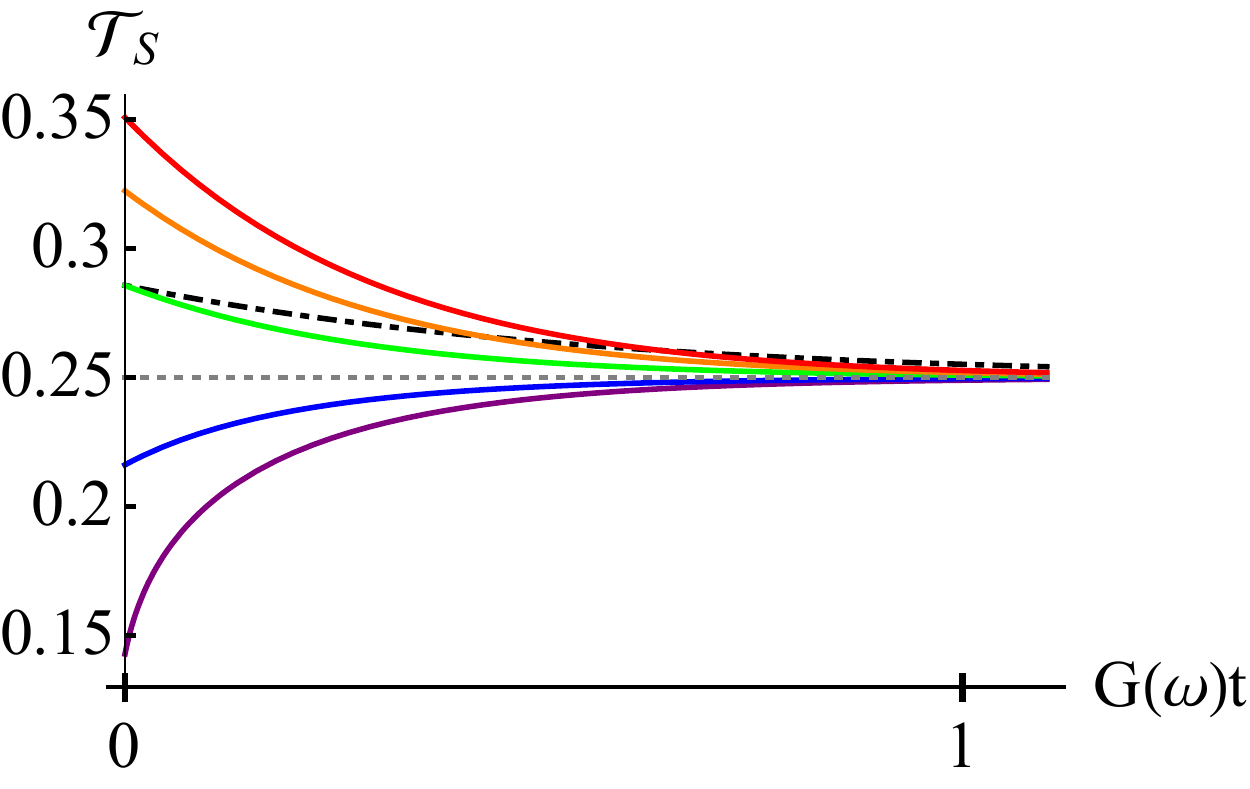}
\caption{(a) Plots of the energy $E_S$ of $S$ as a function of time (normalised by the characteristic evolution time scale, $G(\omega)^{-1}$, see Appendix \ref{esvstime}) for different initial correlations and with $\omega\beta_S=3.5$ and $\omega\beta_B=4$. The purple, blue, green, orange, and red curves corresponds to initial correlations with $\Re \alpha =-e^{-\omega\beta_S}/Z(\beta_S)\simeq-0.028$ (the minimal allowed value), $\Re \alpha=-0.02$, $\Re \alpha=0$, $\Re \alpha=0.015$, and $\Re \alpha =e^{-\omega\beta_S}/Z(\beta_S)\simeq0,028$ (the maximal allowed value), respectively. The associated dotted lines indicate the asymptotic value of each curve, that is the steady state energy of $S$. The dot-dashed black curve is the evolution of the thermal energy (which corresponds to an independent dissipation of each subsystem of $S$). (b) Plots of the apparent temperature ${\cal T}_S$ of $S$ as a function of the (normalised) time for $\omega\beta_S=3.5$ and $\omega\beta_B=4$. Each curve represents different initial correlations identified by the same colour as in (a) and the dot-dashed black curve represents the time evolution of the temperature for an independent dissipation. The dot grey line represents the bath temperature, to which all apparent temperatures eventually converge.
}
\label{enVStime}
\end{figure}

As illustration of the above ideas, Fig. \ref{enVStime} (a) presents plots of $E_S={\rm Tr}_S\rho_S H_S$, the energy of $S$, as a function of time (normalised by the characteristic evolution time scale, $G(\omega)^{-1}$, see Appendix \ref{esvstime}) for values of $\Re\alpha$ ranging from $-e^{-\omega\beta_S}/Z(\beta_S)$ to $e^{-\omega\beta_S}/Z(\beta_S)$, with $\omega\beta_S=3.5$ and $\omega \beta_B=4$. Thus, this is a situation where one would expect the heat to flow from $S$ (the hottest) to $B$ (the coldest). The green curve corresponds to no initial correlations whereas the purple curve and the red curves correspond to the minimal and maximal correlations allowed, namely $\Re\alpha = -e^{-\omega\beta_S}/Z(\beta_S)\simeq -0.028$ and $\Re\alpha=e^{-\omega\beta_S}/Z(\beta_S)\simeq 0.028$, respectively. The blue and orange curves correspond to intermediate values ($\Re\alpha = -0.02$ and $\Re\alpha= 0.015$, respectively). The dotted lines associated to each curve indicate the asymptotic value (i.e. the steady state energy of $S$). The derivation of the expression of $E_S$ as a function of time is detailed in Appendix \ref{esvstime}. 

These curves deserves several comments. First, one can see from the purple and blue curves that not only the heat flow is reversed for these initial correlations, but the steady state of the pair has actually a higher energy than initially (indicated by the dotted coloured lines): $S$ is heated up by a colder bath. In the following, we refer to this kind of heat flow reversal between the initial and steady state as {\it permanent} heat flow reversal. Note that in principle the conditions \eqref{realpha1} and \eqref{realpha2} only guarantee the reversal of the heat flow at the initial instant of times, so that the permanent heat flow reversal observed for the purple and blue curves was not necessarily expected. In Appendix \ref{heatexchange} we come back in more details on this point and show that
the conditions for permanent heat flow reversal are the same as \eqref{condoncor1} and \eqref{condoncor2} when substituting $\alpha_c$ by $\alpha_p$, where $|\alpha_p|>|\alpha_c|$.
In other words, a permanent heat flow reversal requires (slightly) stronger initial correlations than heat flow reversal.
Secondly, the permanent heat flow reversal can be very important (up to $50\%$ of the initial energy). This is the object of the next section ``Maximal reversal of heat exchanges''.
Finally, as a reference, we also plot (dot-dashed black curve) the evolution of the thermal energy which corresponds to the situation where each subsystem interacts independently with $B$ so that it remains in a thermal state at all times and equilibrates to the thermal state at the bath temperature. Comparing the green and dot-dashed black curves one recovers the effect of {\it mitigation} of the bath's action described in \cite{bathinducedcohTLS,bathinducedcoh}. 

It is also insightful to look at the time evolution of the apparent temperature of $S$. In Fig. \ref{enVStime} (b) we plot ${\cal T}_S$ as a function of (the normalised) time for $\omega\beta_S=3.5$ and $\omega\beta_B=4$. The derivation of the time evolution of ${\cal T}_S$ can be found in Appendix \ref{esvstime}. Each curve corresponds to a different initial correlation (characterised by $\Re\alpha$) identified by the same colour as in Fig. \ref{enVStime} (a). One can see the huge impact correlations have on the apparent temperature, which explains the subsequent impact on the energy $E_S$, Fig. \ref{enVStime} (a). Moreover, while reaching the steady state, all curves eventually converge to the bath temperature $1/\beta_B=1/4$ (dotted grey line). The dot-dashed black curve represents the time evolution of the temperature for independent dissipation of each subsystems.

Finally, one can also observe that the dissipation process is slower through independent dissipation (dot-dashed black curve) than collective dissipation (full curves). This phenomenon of equilibration speed-up stemming from collective dissipation was recently studied in \cite{Manatuly_2019,Turkpence_2019}.

\subsection{Maximal heat flow reversal}\label{secmaxhfr}
\begin{figure}
\centering
(a)\includegraphics[width=7cm, height=4.5cm]{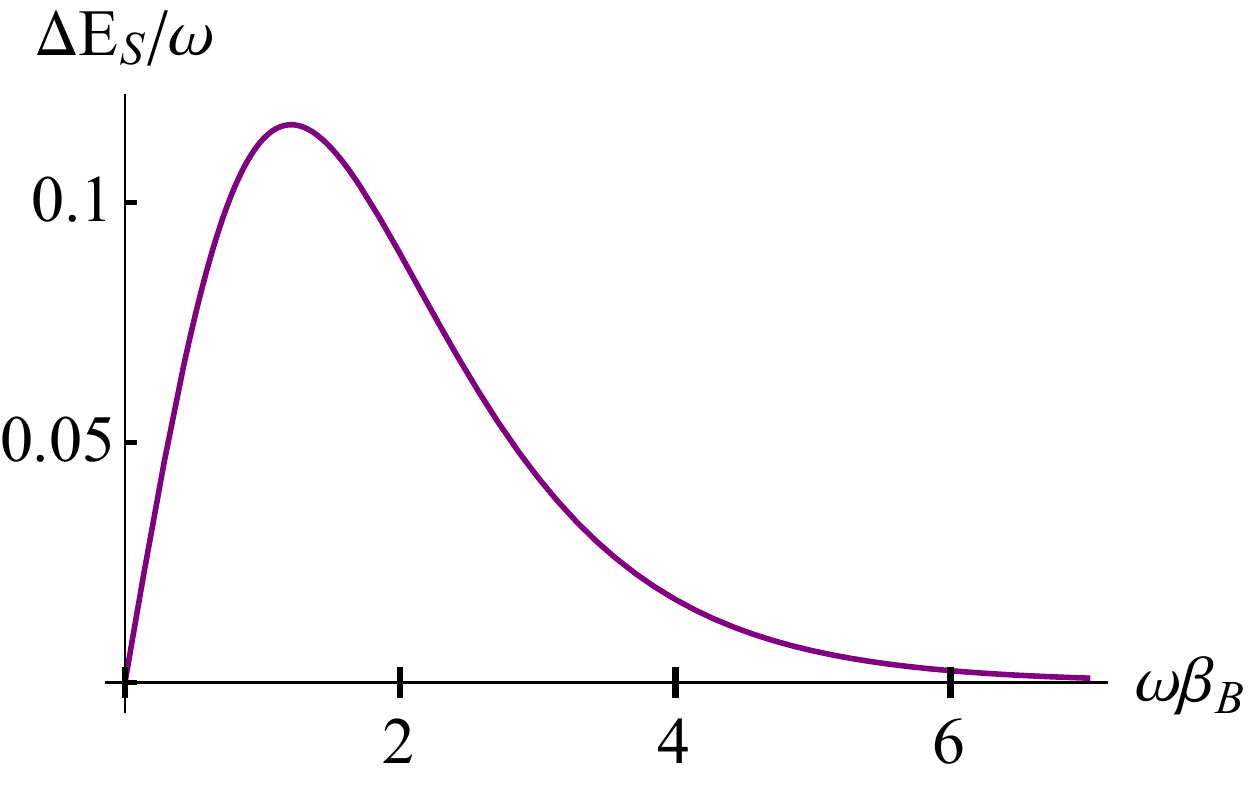}
(b)\includegraphics[width=7cm, height=4.5cm]{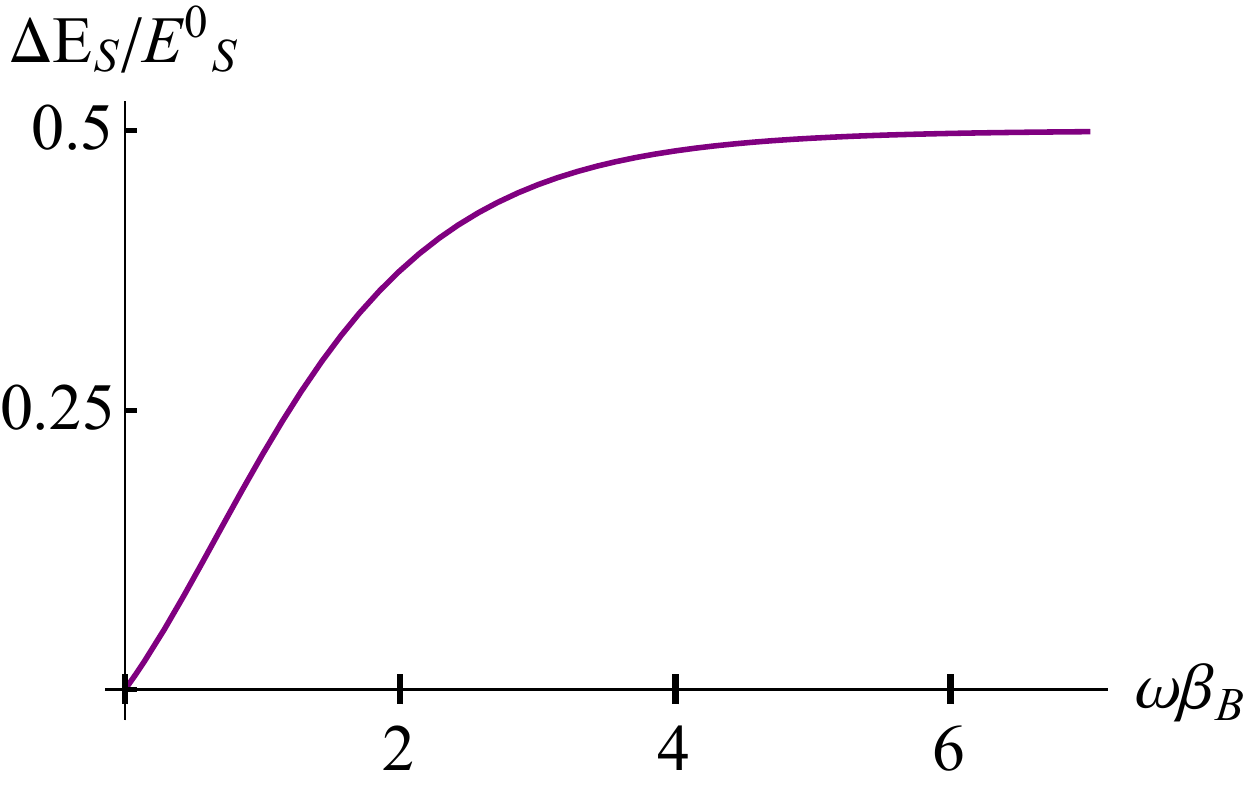}
\caption{(a) Plot of the maximal heat flow reversal $\Delta E_S:=E_S^{\infty}(\beta_B,\beta_B,\Re\alpha_{\rm min})-E_S^0(\beta_S)$ as a function of $\omega\beta_B$, and for $\beta_S=\beta_B$. 
(b) Plot of the maximal heat flow reversal normalised by the initial energy $\Delta E_S/E_S^0(\beta_S)$ as a function of $\omega\beta_B$.
}
\label{reversalVSbb}
\end{figure}

In this section we briefly study how large can be the permanent heat flow reversal with respect to the initial energy of $S$. In Appendix \ref{esvstime} we derive an expression of the steady state energy (see also \cite{bathinducedcohTLS}). We obtain
\be\label{expressioneinfty}
\frac{E_S^{\infty}(\beta_S,\beta_B,\Re\alpha)}{\omega}= 1+\frac{(\Re \alpha + z(\beta_S))\left(e^{-2\omega\beta_B}-1\right)}{1+e^{-\omega\beta_B}+e^{-2\omega\beta_B}}
\ee
with $z(\beta_S):=(1+e^{-\omega\beta_S}+e^{-2\omega\beta_S})/Z(\beta_S)$, recalling that $Z(\beta_S)=(1+e^{-\omega\beta_S})^2$.
For $\beta_B>0$ and $\beta_S<\beta_B$, the maximal heat flow reversal is achieved for an initial correlation with the minimal allowed real part, $\Re \alpha =\Re\alpha_{\rm min}:= -e^{-\omega\beta_S}/Z(\beta_S)$ (implying $\alpha = \Re \alpha$), and for $\beta_S$ tending to $\beta_B$ (as one could expect). In Fig. \ref{reversalVSbb} (a) we show the plot of the maximal heat flow reversal $\Delta E_S:=E_S^{\infty}(\beta_S,\beta_B,\Re\alpha_{\rm min})-E_S^0(\beta_S)$ 
as a function of $\omega\beta_B$ for $\beta_S=\beta_B$, 
 where $E_S^0(\beta_S)$ denotes the initial energy, equal to the thermal energy at inverse temperature $\beta_S$, namely $E_S^0(\beta_S)=2\frac{e^{-\omega\beta_S}}{1+e^{-\omega\beta_S}}$. One can see that the permanent heat flow reversal can go up to an energy equal to $0.12\omega$ and represent a gain of energy up to $50\%$ of the initial energy as shown in Fig. \ref{reversalVSbb} (b). This is a very significative effect. 

For $\beta_S>\beta_B>0$, the maximal heat flow reversal is achieved for an initial correlation with the maximal allowed real part, $\Re \alpha =\Re\alpha_{\rm max}:= e^{-\omega\beta_S}/Z(\beta_S)$ (implying $\alpha = \Re \alpha$), and for $\beta_S$ tending to $\beta_B$. The corresponding plots are the same as Fig. \ref{reversalVSbb} (a) and (b) but with negative signs. Finally, for $\beta_B<0$, the results are similar (and can be obtained from the above ones just by inverting $\Re\alpha_{\rm min}$ and $\Re\alpha_{\rm max}$).

\subsection{Conversion of correlations into energy}\label{secconversion}
In the previous section \ref{secmaxhfr}, we show that maximal (permanent) heat flow reversals happen for extremal values of the initial correlations and for $\beta_S=\beta_B$. 
This situation is also interesting by itself: the on going energy exchanges occur only thanks to the initial correlations. In other words, this is a direct conversion of correlations into energy. 
Driven by the curiosity about this intriguing process, we  mention some of its interesting properties. The following considerations go slightly beyond the scope of this paper, but we found them worth the following brief overview.

\begin{figure}
\centering
(a)\includegraphics[width=7cm, height=4.5cm]{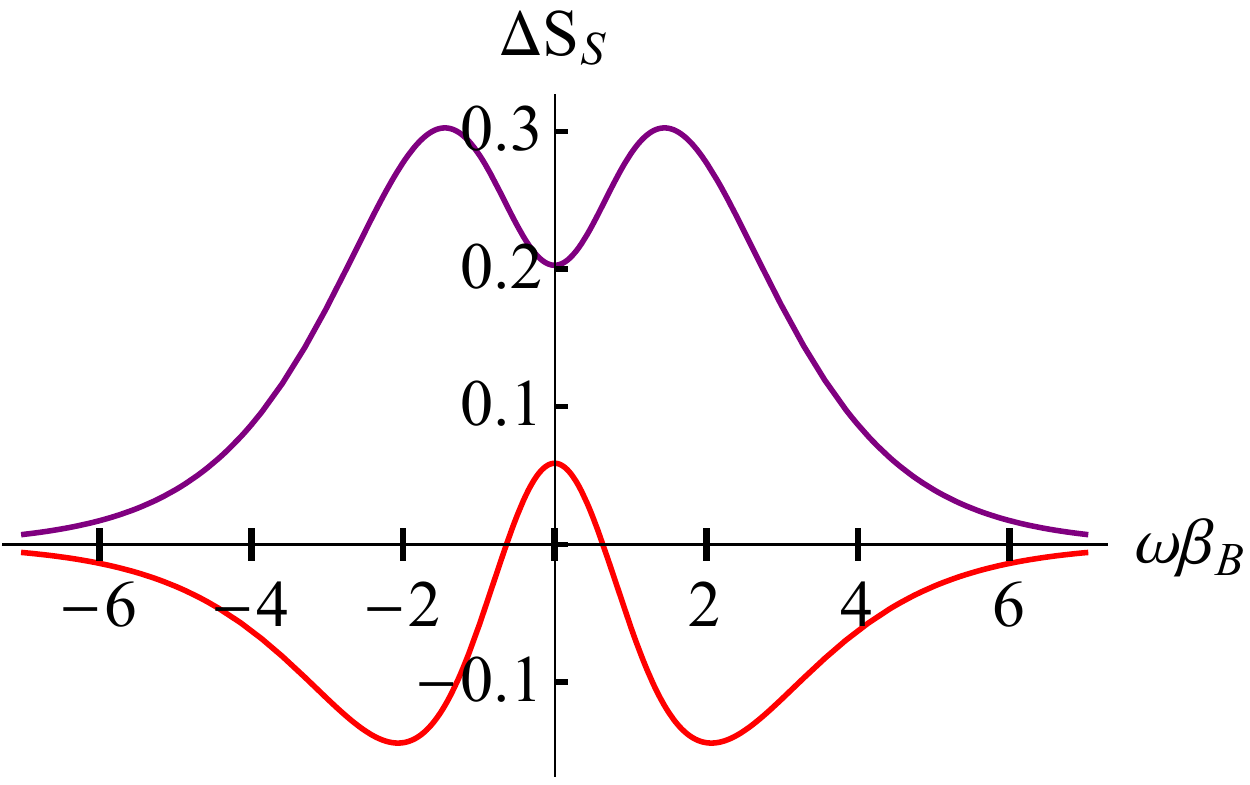}
(b)\includegraphics[width=7cm, height=4.5cm]{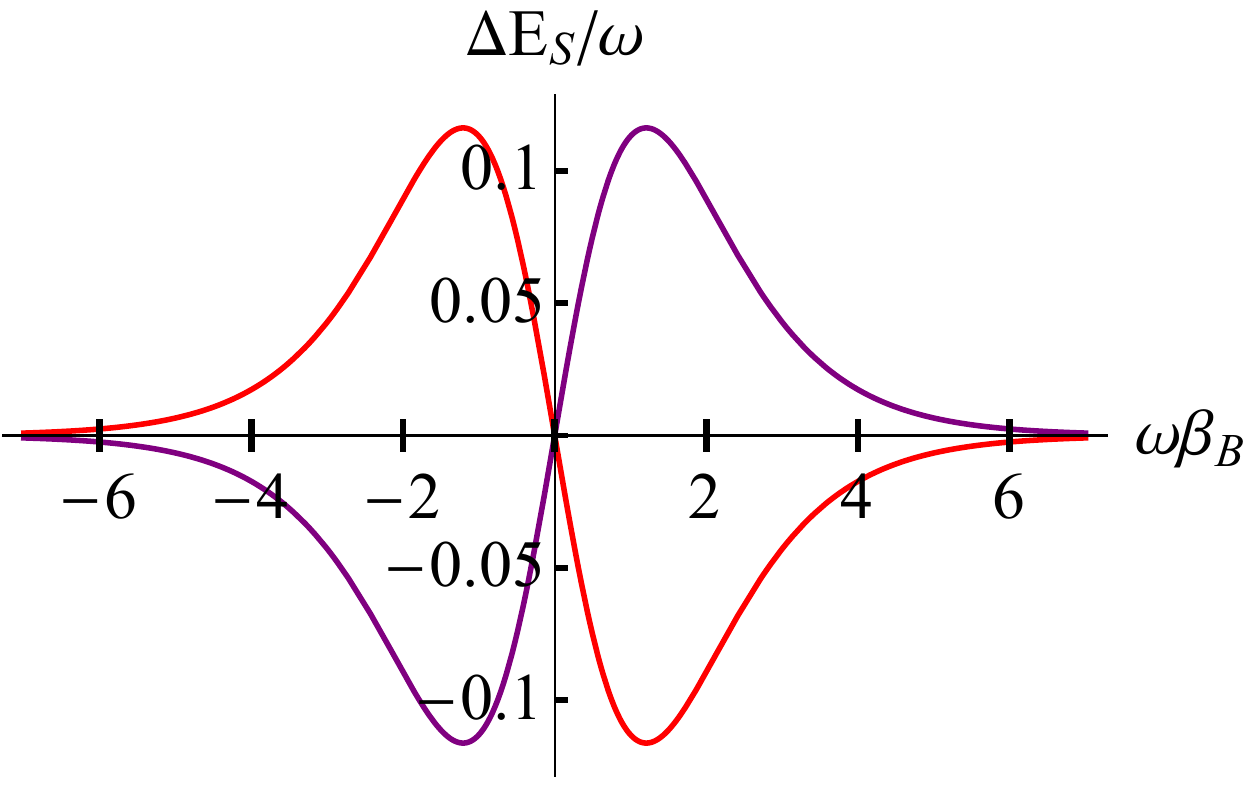}
\caption{Plots of (a) the entropy variation $\Delta S_S=S_S^{\infty}-S_S^0$ between the initial and steady state of $S$, and (b) the energy exchange $\Delta E_S=E_S^{\infty}-E_S^0$, 
as functions of $\omega\beta_B$ for $\Re\alpha =\Re\alpha_{\rm min}$ (purple curves) and $\Re\alpha =\Re\alpha_{\rm max}$ (red curves).
}
\label{propVSbb}
\end{figure}

We first focus on the variation of entropy of $S$ during the conversion process. Using the expression of the steady state derived in Appendix \ref{esvstime} one can derive the expression of $S_S^{\infty}$, the steady state entropy of $S$ (see Appendix \ref{entropyexpression} for the detail of the expression).
 Fig. \ref{propVSbb} (a) displays the plot of the the variation of entropy $\Delta S_S:=S_S^{\infty}-S_S^{0}$ between the initial and steady state as a function of $\omega\beta_B$ for $\alpha=\Re\alpha =\Re\alpha_{\rm min}$ (purple curve) and $\alpha=\Re\alpha =\Re\alpha_{\rm max}$ (red curve).  
  It is insightful to analyse these curves while keeping an eye on Fig. \ref{propVSbb} (b) which displays the energy exchange $\Delta E_S$ as a function of $\omega \beta_B$, still for $\alpha=\Re\alpha =\Re\alpha_{\rm min}$ (purple curve) and $\alpha=\Re\alpha =\Re\alpha_{\rm max}$ (red curve). Note that Fig. \ref{propVSbb} (b) is just an extension of Fig. \ref{reversalVSbb} (a) to negative bath temperature and displaying the symmetric situation of $\alpha=\Re\alpha =\Re\alpha_{\rm max}$.
 
  Interestingly, one can see that the value $\omega\beta_B\simeq 0.7$ is very peculiar for the red curve. Indeed, around this point, the initial and final entropy are the same while the energy variation is highly negative as shown by Fig. \ref{propVSbb} (b). We have here an interesting setup which reproduces perfectly (at least around the value $\omega\beta_B =0.7$) the ideal quantum battery or external (classical) energy source \cite{Kosloff_2013,Kosloff_2014}: it delivers energy without changing its entropy, which corresponds to the idea of perfect work generator. Then, could one say that this process is a conversion of correlations into perfect work? It also rises recurrent questions regarding the nature of the difference between heat and work as it is usually understood that energy exchanges with a {\it thermal} bath are exclusively heat (independently of the sign of the energy exchanges).

\section{Conclusion}
This work presents a surprising phenomenon: indistinguishability and correlations between subsystems can reverse the heat flow between an ensemble and its bath. 
Additionally, in contrast to previous works \cite{Micadei_2017,Partovi_2008, Jennings_2010, Jevtic_2012, Henao_2018}, this does not involve reversal of the arrow of time. This phenomenon is firstly unveiled in a broader context, including several type of degenerate systems, as a consequence of the impact of correlations and non-energetic coherences on the system's apparent temperature. This emphasises that the apparent temperature \cite{paperapptemp} is not merely a mathematical tool, but that it has real physical meaning and consequences which could be tested experimentally relatively easily in pairs of two-level systems.

Then, adopting the formalism used in \cite{Micadei_2017, Jennings_2010}, we provide an alternative view: the degradation of initial internal correlations provides a kind of ``fuel'' for heat flow reversal. Still, there is a fundamental difference with \cite{Micadei_2017, Jennings_2010}: in our protocol, system and bath are initially uncorrelated, which guarantees the positivity of the entropy production and therefore does not affect the arrow of time. Even though, we show that initial {\it internal} correlations (correlations within $S$), can also lead to heat flow reversal.
 In this sense, our setup has the advantage that is does not need previous preparation steps involving both system and bath (not always accessible or controllable), but only preparation steps on the system.
Note also that since the reported effect only requires a degenerate system or correlated subsystems interacting collectively with a bath, it might occur in several many-body experimental setups or even in organic structures. 

Aiming at accessible experimental realisations, we then focus on a pair of two-level systems (used in \cite{Micadei_2017}). We show that the heat flow reversal can be very large, leading to {\it permanent} heat flow reversal up to $50\%$ of the initial energy. One might expect this phenomenon to become larger and larger as the number of subsystems increases in a similar way as indistinguishability has growing impact in spin ensembles of increasing spin number \cite{bathinducedcoh}. It would be indeed interesting to investigate such heat flow reversal in larger systems. 


We finally broaden the study pointing at intriguing processes of correlations-to-energy conversion. In particular, we point out a regime where the correlations-to-energy conversion operates at constant entropy, reproducing the behaviour of external energy source or pure work source. 

Our results uncover the central role played by non-energetic quantum coherences in collective dissipative processes, leading to diverse curious phenomena such as heat flow reversals. This suggests that the heat flow reversal pointed out in this paper has no classical counterpart unlike the one previously known relying on initial correlations between system and bath. An interesting question would be to investigate in more details whether non-energetic coherences also play a special role in entropy production as compared to the role played by energetic coherences recently studied in \cite{Santos_2019}.



\acknowledgements
This  work  is  based  upon  research  supported  by  the
South  African  Research  Chair  Initiative, Grant No. UID 64812 of  the  Department  of  Science  and  Technology of the Republic of South Africa and  National  Research Foundation of the Republic of South Africa.

\appendix
\numberwithin{equation}{section}

\section{Heat flow and apparent temperature}\label{appheatflow}
The reduced dynamics of $S$ is given by the following Markovian master equation (valid under weak coupling)
\bea\label{appme}
\dot{\rho}_t&=&   \Gamma(\omega) \left[{\cal A}\rho_t {\cal A}^{\dag} -{\cal A}^{\dag}{\cal A}\rho_t\right] + {\rm h.c.}\nn\\
&+& \Gamma(-\omega) \left[{\cal A}^{\dag}\rho_t {\cal A} -{\cal A}{\cal A}^{\dag}\rho_t\right] + {\rm h.c.},
\eea
where $\Gamma(\omega)=\int_0^{\infty}ds e^{i\omega s}{\rm Tr}\rho_B B(s)B$, and $B(s)$ is the bath operator $B$ introduced in \eqref{coupling} in the interaction picture (with respect to its free Hamiltonian). 
The heat flow between $S$ and $B$, given by the energy variation $E_S^t:={\rm Tr} \dot\rho_S^t H_S$, can be expressed using the above master equation. Using the commutation relation of the eigenoperator $[{\cal A}^{\dag},H_S]=-\omega{\cal A}^{\dag}$, one can rewrite $\dot E_S$ in the following interesting form,
\bea
\dot E_S  &=&\omega G(\omega)\bra{\cal A}{\cal A}^{\dag}\ket_{\rho_S^t}\big(e^{-\omega/T_B}-e^{-\omega/{\cal T}_S}\big),\nn\\
\eea
where $G(\omega):= \Gamma(\omega)+\Gamma^{*}(\omega)=\int_{-\infty}^{\infty} ds e^{i\omega s}{\rm Tr}\rho_B B(s)B$ and the bath temperature $T_B$ can be defined through $e^{-\omega/T_B}:= G(-\omega)/G(\omega)$ \cite{paperapptemp,Alicki_2014,Alicki_2015}. The apparent temperature \cite{paperapptemp} ${\cal T}_S:= \omega \left(\ln\frac{\bra {\cal A}{\cal A}^{\dag}\ket_{\rho_S}}{\bra{\cal A}^{\dag}{\cal A}\ket_{\rho_S}}\right)^{-1}$, introduced in the main text in \eqref{apptemp}, determines the sign of the heat flow between $S$ and $B$. Indeed, $\dot E_S \geq 0$ ($\dot E_S \leq0$) if and only if ${\cal T}_S \leq 1/\beta_B$ (${\cal T}_S \geq 1/\beta_B$). Thus, ${\cal T}_S$ can be thought as an apparent temperature of $S$, ``seen from the point of view of $B$'', extending the notion of temperature to non-thermal states. Importantly, when $S$ is in a thermal state, the apparent temperature coincides with the usual notion of temperature. One interesting property of the apparent temperature is to take into account contributions from coherences and correlations \cite{paperapptemp}.

\section{Equality of $\bra \ca^{\dag} \ca \ket_{\rm cor} = \bra \ca \ca^{\dag} \ket_{\rm cor}$}\label{equalityofcor}
The equality of $\bra \ca^{\dag} \ca \ket_{\rm cor} = \bra \ca \ca^{\dag} \ket_{\rm cor}$ mentioned in the main text is a direct consequence of the delocalised nature of the correlation term $\chi^0$. Denoting by $a_i$ and $a_i^{\dag}$ the local eigenoperators of each subsystems $S_i$ so that $\ca = \sum_i a_i$ ($\ca^{\dag}=\sum_i a_i^{\dag}$), we have
\bea
\bra \ca^{\dag} \ca \ket_{\rm cor} &=& \sum_{i,j }{\rm Tr} \chi^0 a_ia_j^{\dag} \nn\\
&=& \sum_i {\rm Tr} \chi^0 a_ia_i^{\dag} + \sum_{i\ne j}{\rm Tr} \chi^0 a_ia_j^{\dag} \nn\\
&=&\sum_{i\ne j}{\rm Tr} \chi^0 a_ia_j^{\dag} \nn\\
&=&\sum_{i\ne j}{\rm Tr} \chi^0 a_j^{\dag}a_i \nn\\
&=& {\rm Tr} \chi^0 \ca^{\dag} \ca\nn\\
&=& \bra \ca \ca^{\dag} \ket_{\rm cor}
\eea
where the term ${\rm Tr}\chi^0 a_ia_i^{\dag} = {\rm Tr}\chi^0 a_i^{\dag}a_i ={\rm Tr}_i a_i^{\dag}a_i{\rm Tr}_{S/S_i}(\chi^0) $ is null for all $i$ since ${\rm Tr}_{S/S_i}(\chi^0)$, the partial trace of $\chi^0$ over all subsystems other than $S_i$, is null (the correlations does not contribute to the local states).

\section{Expression of the energy of $S$ as a function of time}\label{esvstime}
The following derivation was already detailed in the Supplementary Material of \cite{paperapptemp} (Section VIII). We just reproduce it here for the sake of completeness. Based on the form of the system-bath coupling \eqref{coupling} one can derived the following master equation, valid under weak coupling, which legitimates the Born and Markov approximations \cite{Petruccione_Book,Cohen_Book},
\bea\label{indsm}
\dot{\rho}_S^I &=& -i\Omega_L \sum_{i=1}^2 [\sigma_i^{+}\sigma_i^{-},\rho_S^I]-i\Omega_{1,2}[\sigma_1^{+}\sigma_{2}^{-}+\sigma_1^{-}\sigma_{2}^{+},\rho_S^I] \nn\\
&&+ G(\omega)(2S^-\rho_S^I S^+ -S^+S^-\rho_S^I - \rho_S^IS^+S^-)\nn\\
&&+ G(-\omega) (2S^+\rho_S^I S^- -S^-S^+\rho_S^I - \rho_S^IS^-S^+).\nn\\
\eea
In the above equation we used the bath spectral density $G(\omega)$ introduced in Appendix \ref{appheatflow}. Furthermore, the operators $\sigma_i^+$ and $\sigma_i^-$ are the ladder operators of the two-level system $S_i$ (spin or atom) defined in Section \ref{secTLS}, $S^+ = \sum_{i=1}^2\sigma_i^+$ and $S^-= \sum_{i=1}^2\sigma_i^- $ are the {\it collective} ladder operators also introduced in Section \ref{secTLS}, $\Omega_L$ is the Lamb shift, and 
the term proportional to $\Omega_{1,2}$ corresponds to the interaction between the two subsystems $S_1$ and $S_2$ (which can be taken to zero if the two subsystems do not interact). The above master equation has the same form as the one describing the dynamics of a pair of two-level atoms interacting with the free space electromagnetic field \cite{Gross_1982}. In such case the interaction $\Omega_{1,2}$ corresponds to the Van der Waals interaction. Note also that the above master equation is valid for any stationary bath \cite{Alicki_2014,Alicki_2015,paperapptemp}, which includes some non-thermal baths. In such situations, the bath inverse temperature $\beta_B$ is substituted by the inverse of the bath apparent temperature which can be simply defined as $\omega[\log[G(\omega)/G(-\omega)]]^{-1}$ \cite{Alicki_2014,paperapptemp}. 

 The dynamics described by \eqref{indsm} can be easily solved by considering the basis $\{|\psi_0\ket,|\psi_+\ket,|\psi_-\ket,|\psi_1\ket\}$ with $|\psi_{\pm}\ket =(|01\ket\pm|10\ket)/\sqrt{2}$, $|\psi_{0}\ket = |00\ket$, and $|\psi_1\ket=|11\ket$. In such basis the collective ladder operators can be expressed as $S^{+}=\sqrt{2}|\psi_+\ket \bra\psi_0|+\sqrt{2}|\psi_1\ket\bra\psi_+|$ and $S^{-}=\sqrt{2}|\psi_0\ket \bra\psi_{+}|+\sqrt{2}|\psi_{+}\ket\bra\psi_1|$. From \eqref{indsm} one obtains the following dynamics for the populations $p_i:=\bra \psi_i|\rho_S|\psi_i\ket$, $i=0,1,+,-$, 
 \bea\label{sys}
 &&\dot{p}_1 = 4 G(-\omega)p_+ - 4G(\omega)p_1\nn\\
 &&\dot{p}_0=4G(\omega)p_+ -4G(-\omega)p_0\nn\\
 &&\dot{p}_+ = 4G(\omega)(p_1-p_+)+4G(-\omega)(p_0-p_+)\nn\\
 &&\dot{p}_- = 0.
 \eea 
 
 The steady state populations can be obtained by canceling the time derivatives in the above system of equations. Alternatively, one can also solve the above system to obtain the time evolution of the populations. This is simplified by noting that $\dot{p}_1+\dot{p}_0+\dot{p}_+ = 0$, which implies that $r:=p_1+p_0+p_+$ is a constant determined by the initial conditions. The system can therefore be reduced to a system of two linearly independent equations (substituting for instance $p_1$ by $r-p_0-p_+$),
\bea
 &&\dot{p}_0=4G(\omega)p_+ -4G(-\omega)p_0\nn\\
 &&\dot{p}_+ = -4[G(\omega)-G(-\omega)]p_0-4[2G(\omega)+G(-\omega)]p_+ \nn\\
 &&\hspace{0.9cm}+ 4G(\omega)r.
 \eea 
The reduced system is diagonalised by the quantities $q^{\pm} := p_+ + (1\pm\sqrt{G(-\omega)/G(\omega)})p_0$, with the associated eigenvalues $a^{\pm}:= 4[ \pm \sqrt{G(\omega)G(-\omega)}-G(\omega)-G(-\omega)]$, so that 
\be 
\dot q^{\pm} = a^{\pm} q^{\pm} + 4G(\omega)r,
\ee
and 
\be
q^{\pm}(t)= e^{a^{\pm} t} q^{\pm}(0) +4G(\omega)r\frac{e^{a^{\pm}t}-1}{a^{\pm}}.
\ee
From the time evolution of $q^{\pm}(t)$ one can deduce straightforwardly the expression for the time evolution of the populations $p_0$, $p_+$, $p_1$,
the energy $E_{S}/\omega = 2p_1+p_{+}+p_{-}$ and the apparent temperature ${\cal T}_{S} = \omega \left(\log\frac{p_0+p_+}{p_1+p_+}\right)^{-1}$. Applying the deduced expressions for different values of the initial correlations ($\Re\alpha$ varying from the minimal to the maximal allowed value), one obtains the plots of Fig. \ref{enVStime}. 

 The corresponding steady state populations are 
 \bea
 p_0^{\infty}&=&r\frac{1}{1 +e^{-\omega\beta_B}+e^{-2\omega\beta_B}},\nn\\
  p_+^{\infty}&=&r\frac{e^{-\omega \beta_B}}{1 +e^{-\omega\beta_B}+e^{-2\omega\beta_B}},\nn\\
  p_1^{\infty}&=& r\frac{e^{-2\omega \beta_B}}{1 +e^{-\omega\beta_B}+e^{-2\omega\beta_B}},\nn\\ 
  p_{-}^{\infty}&=&1-r,
  \eea
  which yields a steady state energy equal to
  \be\label{appssenergy}
  E^{\infty}_S= 1+r\frac{e^{-2\omega\beta_B}-1}{1+e^{-\omega\beta_B}+e^{-2\omega\beta_B}}
    \ee
When the initial state is of the form \eqref{initialstt}, the constant $r$ is equal to 
\be
r=\Re\alpha + z(\beta_S),
\ee
 where $z(\beta_S) := (1+e^{-\omega\beta_S}+e^{-2\omega\beta_S})/Z(\beta_S)$, which leads for the steady state energy to the expression \eqref{expressioneinfty} announced in the main text.

For the coherences, defined as $\rho_{ij}:=\bra \psi_i|\rho_S^I|\psi_j\ket$, $i,j \in \{0,1,+,-\}$, one obtains (including the Lamb shift in the interaction picture), 
\bea 
&&\dot{\rho}_{+,-} = -2\big[G(\omega)+G(-\omega) + i\Omega_{I} \big] \rho_{+,-}\nn\\
&&\dot{\rho}_{1,-} = -\big[2G(\omega)+i\Omega_{I}\big]\rho_{1,-} \nn\\
&&\dot{\rho}_{0,-} = -\big[2G(-\omega)+i\Omega_{I}\big]\rho_{0,-} \nn\\
&&\dot{\rho}_{1,0} = -2[G(\omega)+G(-\omega)]\rho_{1,0}
\eea
which straightforwardly gives $0$ as steady state solution. The dynamics of the two remaining coherences is coupled,
\bea
&&\dot{\rho}_{1,+} = -\big[2(2G(\omega)+G(-\omega))-i\Omega_{I}\big]\rho_{1,+} + 4G(-\omega)\rho_{+,0}\nn\\
&&\dot{\rho}_{+,0} =  -\big[2g(G(\omega)+2G(-\omega))+i\Omega_{I}\big]\rho_{+,0} + 4G(\omega)\rho_{1,+},\nn\\
\eea
and also leads to $0$ as steady state solution. Finally, one can write the steady state in the form,
\bea\label{appsteadystate}
  \rho_{S}^{\infty} &=& \frac{r}{1+e^{-\omega\beta_B}+e^{-2\omega\beta_B}}\Big(e^{-2\omega\beta_B}|\psi_1\ket\bra \psi_1|\nn\\
  &&+e^{-\omega\beta_B}|\psi_+\ket\bra\psi_+| + |\psi_0\ket\bra\psi_0|\Big) + (1-r)|\psi_{-}\ket\bra\psi_{-}|.\nn\\
\eea

\section{Permanent heat flow reversal}\label{heatexchange}
As commented in Section \ref{secTLS} in the main text, the conditions \eqref{condoncor1} and \eqref{condoncor2} correspond to reversal of the heat flow at initial times, but does not guarantee that the heat flow remains inverted throughout times until $S$ reaches its steady state, what we call {\it permanent} reversal of the heat flow. We show now that a permanent reversal requires indeed a higher level of initial correlations. From the expression of the steady state energy \eqref{appssenergy} (or \eqref{expressioneinfty} in the main text)
\be
E^{\infty}_S=1+\frac{(\Re\alpha + z(\beta_S))(e^{-2\omega\beta}-1)}{1+e^{-\omega\beta_B}+e^{-2\omega\beta_B}},
\ee
compared to the initial energy $E_S^0:={\rm Tr}\rho^0_SH_S = 2\frac{e^{-\omega\beta_S}}{(1+e^{\-\omega\beta_S}}$, one obtains that a permanent heat flow reversal takes place if and only if
\bea\label{alphap}
\Re\alpha > \alpha_p >0, {~\rm for ~ \beta_S/\beta_B>1}\nn\\
\Re\alpha <\alpha_p <0, {~\rm for ~ \beta_S/\beta_B<1}
\eea
with $\alpha_p:=z(\beta_B)\frac{1+e^{-\omega\beta_B}}{1-e^{-\omega\beta_B}}\frac{1-e^{-\omega\beta_S}}{1+e^{-\omega\beta_S}}-z(\beta_S)$. Note that the above conditions are valid for $\beta_S$ and $\beta_B$ of arbitrary sign. 
A comparison with $\alpha_c$ the critical value required for $\Re\alpha$ to induce heat flow reversal introduced in \eqref{condoncor1} and \eqref{condoncor2} shows that $|\alpha_c|<|\alpha_p|$ which implies that permanent heat flow reversal always requires a higher level of correlations/coherences.

\section{Expression of the steady state entropy}\label{entropyexpression}
The steady state entropy (von Neumann entropy) can be obtained from the expression of the steady state \eqref{appsteadystate}. Since the expression is given in a basis that diagonalises $\rho_S^{\infty}$, it is straightforward to show that the steady state entropy is 
\bea
S^{\infty}_S &:=& -{\rm Tr}\rho_S^{\infty}\ln \rho_S^{\infty}\nn\\
&=& r\ln [(1+e^{-\omega\beta_B}+e^{-2\omega\beta_B})/r ] -(1-r)\ln (1-r)\nn\\
&& +r\omega\beta_B\frac{e^{-\omega\beta_B}+e^{-2\omega\beta_B}}{1+e^{-\omega\beta_B}+e^{-2\omega\beta_B}},
\eea 
where $r:=\bra \psi_0|\rho_S^0|\psi_S\ket+\bra \psi_0|\rho_S^0|\psi_0\ket+\bra \psi_S^0|\rho_0|\psi_0\ket$ is a constant introduced in Appendix \ref{esvstime} together with the basis $\{|\psi_0\ket,|\psi_+\ket,|\psi_-\ket,|\psi_1\ket\}$ defined by $|\psi_{\pm}\ket =(|01\ket\pm|10\ket)/\sqrt{2}$, $|\psi_{0}\ket = |00\ket$, and $|\psi_1\ket=|11\ket$.
For initial states of the form \eqref{initialstt} with $\chi^0$ given by \eqref{corrtermTLS} one has simply $r=z(\beta_S) +\Re\alpha$, with $z(\beta_S):=(1+e^{-\omega\beta_S}+e^{-2\omega\beta_S})/Z(\beta_S)$. Such initial states can be re-written in a diagonal form as 
\bea\label{appssentropy}
\rho_S^0 &=& \frac{e^{-2\omega\beta_B}}{Z(\beta_B)} |\psi_1\ket\bra\psi_1| + \left(\frac{e^{-\omega\beta_B}}{Z(\beta_B)} +|\alpha|\right)|\psi_{+,\phi}\ket\bra\psi_{+,\phi}| \nn\\
&&+\left(\frac{e^{-\omega\beta_B}}{Z(\beta_B)} -|\alpha|\right)|\psi_{-,\phi}\ket\bra\psi_{-,\phi}|+\frac{1}{Z(\beta_B)} |\psi_0\ket\bra\psi_0| \nn\\
\eea
in the basis $\{|\psi_0\ket,|\psi_+\ket,|\psi_-\ket,|\psi_1\ket\}$ defined by $|\psi_{\pm,\phi}\ket =(e^{i\phi}|01\ket\pm e^{-i\phi}|10\ket)/\sqrt{2}$, $|\psi_{0}\ket = |00\ket$, $|\psi_1\ket=|11\ket$, and the phase $\phi:=\arg \alpha$ is the argument of the correlations. Then, the initial entropy can be obtained straightforwardly,
\bea\label{appinitialentropy}
S_S^0&:=&-{\rm Tr} \rho_S^0\ln \rho_s^0 \nn\\
&=& \ln Z(\beta_S) +2\omega\beta_S \frac{e^{-\omega\beta_S}}{1+e^{-\omega\beta_S}}\nn\\
&&-\left(\frac{e^{-\omega\beta_S}}{Z(\beta_S)} +|\alpha|\right)\ln \left[1+ |\alpha|e^{\omega\beta_S}Z(\beta_S)\right]\nn\\
&&-\left(\frac{e^{-\omega\beta_S}}{Z(\beta_S)} -|\alpha|\right)\ln \left[1-|\alpha|e^{\omega\beta_S}Z(\beta_S) \right].\nn\\
\eea
Combining \eqref{appssentropy} and \eqref{appinitialentropy} one can obtain an expression of the variation of entropy $\Delta S_S:=S_S^{\infty}-S_S^0$ which was used to plot the graph \ref{propVSbb} (a).

\end{document}